\documentclass[%
mathleft,%
]{an}
\usepackage{graphicx}
\usepackage{times}
\usepackage{amssymb,amsmath}
\begin{document}

\Pagespan{1}{}
\Yearpublication{2007}%
\Yearsubmission{2007}%
\Month{11}%
\Volume{999}%
\Issue{88}%

\title{Spectral lag of gamma-ray burst caused by the intrinsic spectral evolution and the curvature effect}

\author{Z. Y. Peng \inst{1} \and Y. Yin\inst{2} \and X. W. Bi\inst{3} \and Y. Y. Bao\inst{4}
\and  L. Ma\inst{1}\fnmsep\thanks{Corresponding author:
  \email{astromali@126.com}} }
\authorrunning{Z. Y. Peng et al.}
\institute{Department of Physics, Yunnan Normal University, Kunming,
Yunnan 650092, China \and Department of Physics, Liupanshui Normal
College, Liupanshui 553004, China \and Department of Physics, Honghe
University, Mengzi 661100, China \and Department of Physics, Yuxi
Normal College, Yuxi 653100, China}

\received{} \accepted{} \publonline{}


\abstract {Assuming an intrinsic `Band' shape spectrum and an
intrinsic energy-independent emission profile we have investigated
the connection between the evolution of the rest-frame spectral
parameters and the spectral lags measured in gamma-ray burst (GRB)
pulses by using a pulse model. We first focus our attention on the
evolution of the peak energy, $E_{0,p}$, and neglect the effect of
the curvature effect. It is found that the evolution of $E_{0,p}$
alone can produce the observed lags. When $E_{0,p}$ varies from hard
to soft only the positive lags can be observed. The negative lags
would occur in the case of $E_{0,p}$ varying from soft to hard. When
the evolution of $E_{0,p}$ and the low-energy spectral index
$\alpha_{0}$ varying from soft to hard then to soft we can find the
aforesaid two sorts of lags. We then examine the combined case of
the spectral evolution and the curvature effect of fireball and find
the observed spectral lags would increase. A sample including 15
single pulses whose spectral evolution follows hard to soft has been
investigated. All the lags of these pulses are positive, which is in
good agreement with our theoretical predictions. Our analysis shows
that only the intrinsic spectral evolution can produce the spectral
lags and the observed lags should be contributed by the intrinsic
spectral evolution and the curvature effect. But it is still unclear
what cause the spectral evolution.}

\maketitle

\section{Introduction}
The phenomenon of the observed spectral lag of Gamma-ray burst (GRB)
is very common. Since Cheng et al. (1995) first analyzed the
spectral lags of GRBs based on BATSE Large Area Detector channel 1
(25-50 keV) and channel 3 (100-300 keV) light curves, several
authors have carried out more analysis work on the GRB lags (e.g.,
Norris et al., 2000; Wu \& Fenimore 2000; Chen et al. 2005; Yi et
al. 2006; Peng et al. 2007).

Several attempts to explain the origin of the spectral lag have been
put forward. The dominant behavior observed in GRBs, positive lags,
is that the soft radiation lagging the hard, can be produced by
several reasons. An intrinsic cooling of the radiating electrons
cause the radiation to dominate at lower energies. A Compton
reflection of a medium at a sufficient distance from the initial
hard source will also cause a positive lag. While the photons emit
at relativistic speeds from different latitudes of a convex surface
(the curvature effect) will cause the radiation to be delayed and
softened, thus producing a observed positive lag (e.g., Qin 2002;
Ryde \& Petrosian 2002; Qin et al. 2004; Shen et al. 2005). The
negative lag is defined as the hard radiation lagging behind the
soft. However, few attempts are made to explain the origin of the
negative lag. The possible scenario is a hot medium, say a lepton
cloud surrounding a cooler emitter, which will cause the soft
radiation to be upscattered by inverse Comptonization. The photons
get harder the more scatterings they suffer and thus they are more
delayed. A definitive explanation of the origin of them has not yet
been given. The origin of the spectral lags are still unclear.

Kocevski \& Liang (2003) pointed out that the observed lag is the
direct result of spectral evolution. They measured the rate of
$E_{p}$ decay ($\Phi_{0}$) for a sample of clean single-peaked
bursts with measured lag and used this data to provide an empirical
relation that expresses the GRB lag as a function of the burst's
spectral evolution rate. While the studies of GRB spectra have shown
that GRB spectral evolution is a common phenomenon (e.g.,, Kargatis
et al. 1994; crider et al. 1997; Band 1997; Kaneko et al. 2006;
Butler \& Kocevski 2007).

Recently, Shen et al. (2005) tentatively studied the contribution of
curvature effect of fireball to the lag, and the resulting lags are
very closed to the observed one. Lu et al. (2006) also studied the
spectral lags more detailed based on Doppler effect of fireball (or
in some papers, the curvature effect) (see, Qin 2002; Qin et al.
2004; Peng et al. 2006 for more detailed description) and confirmed
that the curvature effect can produce the observed spectral lags.
They performed more precise calculation with both formulae presented
by Shen et al. (2005) and Qin et al. (2004). Some conclusions
obtained by Shen et al. (2005) were ascertained, and more complete
conclusions on spectral lags resulting from curvature effect were
obtained. Lu et al. (2006) argued that, as long as the whole
fireball surface is concerned, both formulas were identical in the
case of ultra-relativistic motions. Other cases did not study by
Shen et al. (2005) were investigated by Lu et al. (2006) and some
new conclusions were drawn. These seem show that the curvature
effect can indeed produce the spectral lags. In addition, several
studies show that the curvature effect also plays an important role
in the early X-ray afterglow of GRBs where the so-called softening
phenomenon is observed (Qin 2008a, 2008b; Qin 2009).

As pointed out above that the fundamental origin of the observed lag
is the evolution of the GRB spectra and the curvature effect can
produce the observed lags. Moreover it is mentioned clearly that
whilst the curvature effect can account for the main part of the
hardness ratio curves of the GRBs concerned and an intrinsic
spectral evolution is required to fully explain the observed data
(Qin et al. 2006). These motivate our investigations below. How the
influences of intrinsic spectral evolution on the spectral lags are
when the curvature effect plays an important role or does not? Can
the negative lags be produced in the course of intrinsic spectral
evolution? We introduce the theoretical formula of the GRB pulse in
Section 2. Various possible cases of producing the spectral lags
when neglect the curvature effect are studied in Section 3. In
Section 4 we investigate the lags when spectral evolution and
curvature effect take effect simultaneously. We employ a sample to
test the prediction in Section 5. In the last section, we give the
conclusions and discussion.

\section{Theoretical formula}
The observed gamma-ray pulses are believed to be produced in a
relativistically expanding and collimated fireball because of the
large energies and the short time-scales involved. To account for
the observed pulses and spectra of bursts, the Doppler effect (or
curvature effect in some paper) over the whole fireball surface
would play an important role (e.g., Meszaros and Rees 1998; Hailey
et al. 1999; Qin 2002; Qin et al. 2004). The Doppler effect is the
photons emitted from the region on the line of sight and those off
the line of sight an angle of $\theta$ are Doppler-boosted by
different factors and travel different distances to the observer.
Therefore, we adopt the model derived by Qin (2002) and Qin et al.
(2004) to explore the spectral lags. But our attention is
concentrated on the case of $\theta \simeq$0. The count rate within
energy channel $[\nu_{1}, \nu_{2}]$ is rewritten as formula (1).
\begin{equation}
\begin{array}{l}
C(\tau )= [2\pi R_c^3 \int_{\widetilde{\tau }_{\theta ,\min
}}^{\widetilde{\tau } _{\theta ,\max }}\widetilde{I}(\tau
_\theta)(1+\beta \tau _\theta
)^2(1-\tau +\tau _\theta )d\tau _\theta \\
 \times \int_{\nu _1}^{\nu _2}\frac{g_{0,\nu
}(\nu_{0,\theta })}\nu d\nu] \times[hcD^2\Gamma
^3(1-\beta)^2(1+\frac \beta{1-\beta}\tau )^2]^{-1}
\end{array}
\end{equation}
where the $\tau_{\theta}$ is dimensionless relative local time
defined by $\tau _\theta \equiv c(t_\theta -t_c)/R_c$, $t_\theta$ is
the emission time in the observer frame, called local time, of
photons emitted from the concerned differential surface $ds_\theta $
of the fireball ($\theta$ is the angle to the line of sight), $t_c$
is the initial local time which could be assigned to any values of
$t_\theta$, and $R_c$ is the radius of the fireball measured at
$t_\theta=t_c$. Variable $\tau$ is a dimensionless relative
observation time defined by $ \tau \equiv [c(t-t_c)-D+R_c]/R_c$,
where $D$ is the distance of the fireball to the observer, and $t$
is the observation time measured by the distant observer.

In formula (1), $\widetilde{I}(\tau _\theta )$ represents the
development of the intensity magnitude of radiation in the observer
frame, called as a local pulse function, and $g_{0,\nu }(\nu
_{0,\theta })$ describes the rest-frame radiation mechanisms.

For the sake of simplicity we first adopt Gaussian pulse as
rest-frame local pulse. As for the rest-frame radiation spectrum we
employ the most frequently used Band function (Band et al. 1993)
since it is rather successfully employed to fit the spectra of the
GRBs. The two forms are rewritten as formulas (2) and (3).
\begin{equation} \widetilde{I}(\tau _\theta
)=I_0\exp [-(\frac{\tau _\theta -\tau _{\theta ,0} }\sigma )^2]
\qquad (\tau _{\theta ,\min }\leq \tau _\theta ),
\end{equation}

\begin{equation}
g_{0,\nu}(\nu _{0,\theta })=
  \left\{
   \begin{array}{c}
 (\frac{\nu
_{0,\theta }}{\nu _{0,p}})^{1+\alpha _0}\exp [-(2+\alpha _0)\frac{
\nu _{0,\theta }}{\nu _{0,p}}],  \\
\qquad\quad (\frac{\nu  
_{0,\theta }}{\nu _{0,p}}<
\frac{\alpha _0-\beta _0}{2+\alpha _0}) \\
(\frac{\alpha _0-\beta _0}{2+\alpha _0})^{\alpha _0-\beta _0}\exp
(\beta_0-\alpha _0)(\frac{\nu _{0,\theta }}{\nu _{0,p}})^{1+\beta
_0}\\ \qquad(\frac{\nu _{0,\theta }}{\nu _{0,p}}\geq\frac{\alpha
_0-\beta _0}{2+\alpha _0}),  \\
   \end{array}
  \right.
  \end{equation}
where $I_0$, $\sigma$ and $\tau_{\theta,min}$ are constants,
$\alpha_0$ and  $\beta_0$ are the low- and high-energy indices in
the rest frame, respectively, and $\nu_{0,p}$ is the rest-frame peak
frequency.

Light curves determined by formula (1) are dependent on the integral
limits $\widetilde{\tau }_{\theta ,\min }$ and $\widetilde{\tau
}_{\theta ,\max }$, which are determined by the concerned area of
the fireball surface, together with the emission ranges of the
radiated frequency and the local time. The integral limits are only
determined by
\begin{equation}
\left\{
   \begin{array}{c}
\widetilde{\tau }_{\theta ,\min }=\max\{\tau _{\theta,\min
},{\frac{\tau-1+\cos \theta _{\max }}{1-\beta \cos \theta _{\max }}}\}\\
\widetilde{\tau }_{\theta ,\max }=\min \{\tau _{\theta ,\max
},\frac{\tau -1+\cos \theta _{\min }}{(1-\beta \cos \theta _{\min
}}\},
\end{array}
  \right.
\end{equation}
where $\tau_{\theta, \min}$ and $\tau_{\theta, \max}$ are the lower
and upper limit of $\tau_{\theta}$ confining $\widetilde{I}(\tau
_\theta )$, and $\theta_{\min}$ and $\theta_{\max}$ are confined by
the concerned area of the fireball surface. The radiations are
observable within the range of
\begin{equation}
   \begin{array}{c}
(1-\cos \theta _{\min })+(1-\beta \cos \theta _{\min })\tau _{\theta
,\min } \leq \tau \leq \\
(1-\cos \theta _{\max })+(1-\beta \cos \theta _{\max })\tau _{\theta
,\max }
\end{array}
\end{equation}

Due to the constraint to the lower limit of $\tau_{\theta}$, which
is $\tau_{\theta,min}>-1/\beta$ (see, Qin et al. 2004), we assign
$\tau_{\theta,0} = 10\sigma+\tau_{\theta,min}$ so that the interval
between $\tau_{\theta,min}$ and $\tau_{\theta,0}$ would be large
enough to make the rising phase of the local pulse close to that of
the Gaussian pulse. From formula (2) one can obtain
\begin{equation}
\Delta\tau_{\theta,FWHM} = 2\sqrt{ln 2}\sigma,
\end{equation}
which leads to
\begin{equation}
\sigma=\Delta\tau_{\theta,FWHM}/2\sqrt{ln 2},
\end{equation}
where $\Delta\tau_{\theta,FWHM}$ is the $FWHM$ (full width at half
maximum) of the Gaussian pulse. From the relation between
$\tau_{\theta}$ and $t_{\theta}$, one gets
\begin{equation}
\Delta t_{\theta,FWHM}=(R_c/c)\Delta \tau_{\theta, FWHM}.
\end{equation}
In the following sections, we assign $(2\pi R_{c} ^{3}
I_{0})/(hcD^{2})$=1, $R_c=3\times10^{15}$ cm, $\tau_{\theta,min}$= 0
.

Because peak times of different light curves associated with
different frequency intervals $[\nu_1, \nu_2]$ or different energy
bands $[E_1, E_2]$ are different, following Shen et al. (2005) and
Lu et al. (2006), we define the spectral lags as the differences
between the pulse peak times in two different energy channels, a
lower energy channel $[E_1, E_2]$ and a higher energy channel $[E_3,
E_4]$. In the following analysis we only consider the energy channel
pair: BATSE channels 1 (25--50 keV) and 3 (100--300 keV) which we
denote the lag as $lag_{13}$ since it was investigated by most of
authors (e.g., Norris et al. 2000; Chen et al. 2005).

\section{Spectral lags caused by the spectral evolution alone}
Since Kocevski \& Liang (2003) pointed out that the observed lag is
the result of spectral evolution, let us first investigate the
spectral lags due to the spectral evolution alone. That is we
consider the radiations emitted from a very small area with
$\theta_{min}=0$ and $\theta_{max} =10 ^{-5} \times 1/\Gamma$ where
the photons emitted travel almost same distances to the observer. In
this case we think that the curvature effect has no role on the
producing of spectral lag.

Observations show that the spectral parameters of the GRBs vary with
time. In addition, the parameters of the Band function low- and
high-energy indices $\alpha$, $\beta$ and the peak energy of GRB's
$\nu F_{\nu}$ spectrum $E_{p}$ in the observed frame are mainly
distributed within $-1.5$
--- $-0.5$, $-3.2$ --- $-2.1$ and 100 --- 700 keV and their typical
values are $\alpha=-1$ and $\beta=-2.25$, $E_{p}=250$ keV,
respectively (Preece et al. 1998, 2000; Kaneko et al. 2006). Qin
(2002) showed the Doppler effect can not affect the intrinsic
spectral shapes. When we take the Doppler effect of fireballs into
account, the observed peak energy would be related to the peak
energy of the rest-frame Band function spectrum ($\alpha_{0}
=\alpha$, $\beta_{0}=\beta$) by $E_{p}\simeq1.67\Gamma E_{0,p}$
(see, Table 4 in Qin 2002). In the following we first consider the
common evolutionary model of the rest-frame spectral parameters to
investigate the spectral lags. Then the other possible cases are
also taken into account.

\subsection{Spectral lags of the rest-frame spectral parameters remain constant}
We first explore if the spectral lags would occur when the
rest-frame spectral parameters remain unchanged to check the effect
of spectral evolution on the spectral lag. The typical values of the
low- and high-energy indices $\alpha_0=-1$ and $\beta_0=-2.25$ and
$E_{0,p}=250/1.67 \Gamma$ keV are taken. Here we assign the
$\sigma=1.0$ and take $\Gamma$ = 100, 200, 300, 400, 500, 600, 700,
800, 900, 1000, respectively since the Lorentz factor of fireball
are generally larger than 100. We find that all the spectral lags
$lag_{13}$ in this case are 0. This suggests that there has no
spectral lags when the rest-frame spectral parameters are constant
and the curvature effect does not play a role on the spectral lag.
Fig. 1 (panel (a)) gives a example plot for this case. The lines of
the peak times of the two pulses are superposed each other.

\begin{figure*}
\centering \resizebox{2in}{!}{\includegraphics{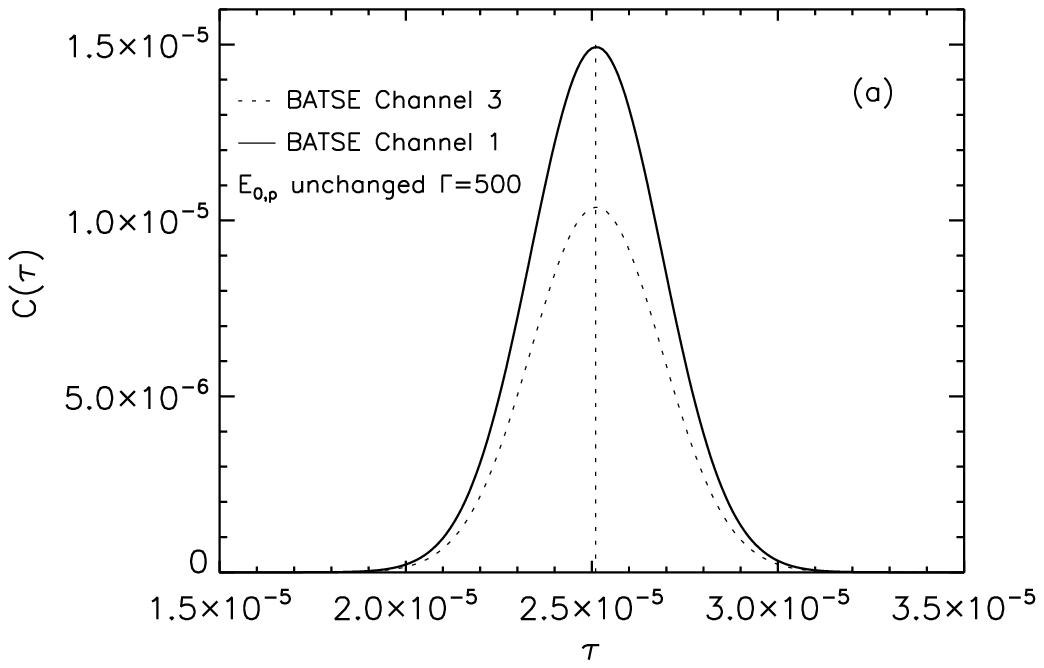}}
\resizebox{2in}{!}{\includegraphics{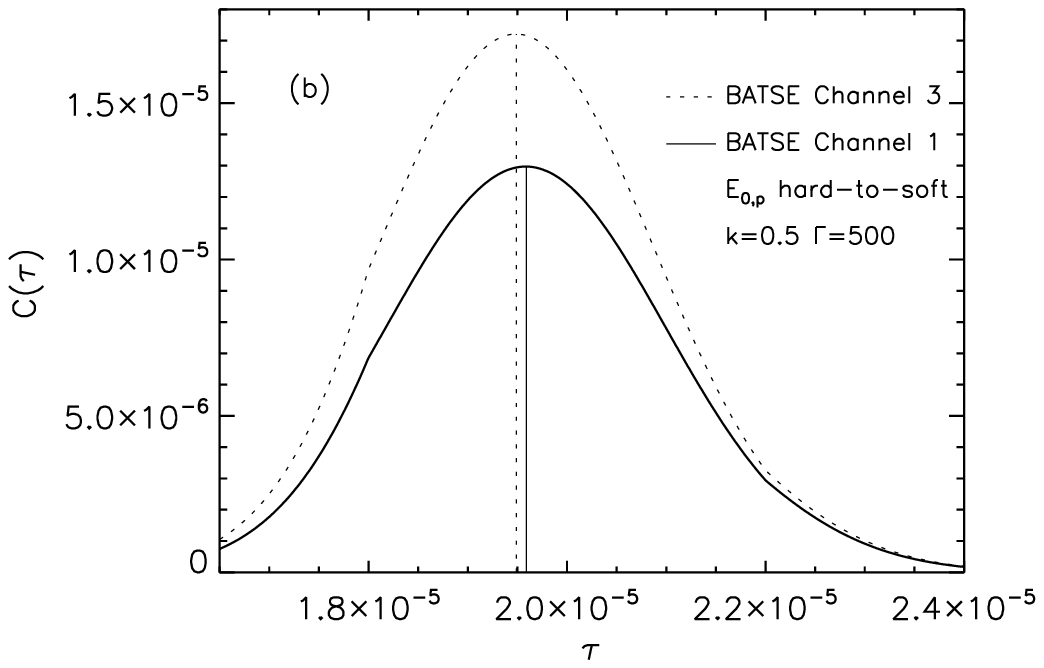}}
\resizebox{2in}{!}{\includegraphics{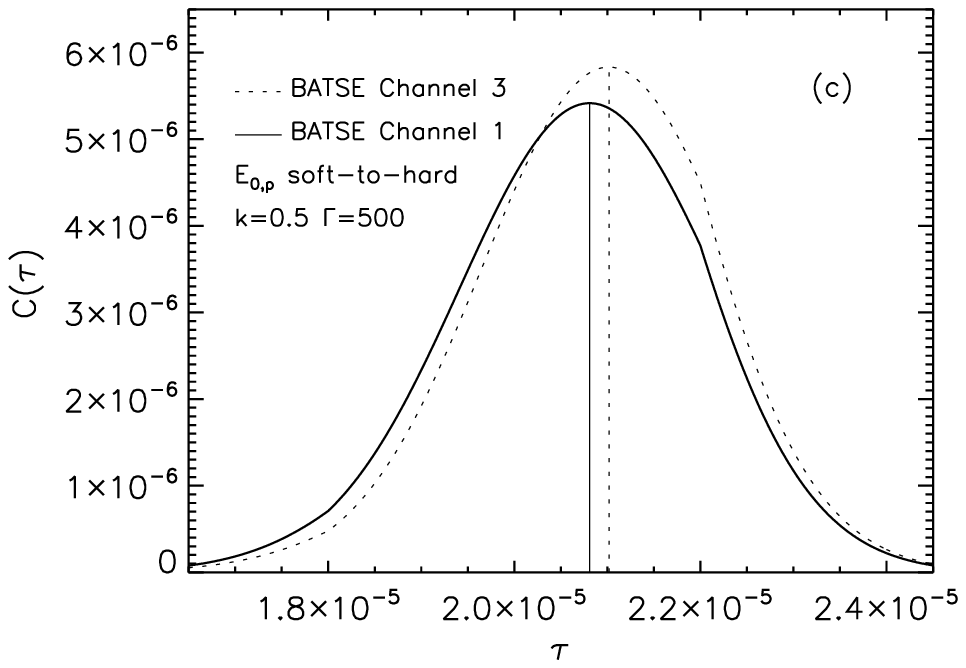}} \caption{Example light
curves of the Band spectral parameters $E_{0,p}$ remain unchanged
(a), the $E_{0,p}$ varying from hard to soft (b) and soft to hard
(c), where the solid and dashed curves represent the light curves of
BATSE channels 1 and 3, respectively.} \label{}
\end{figure*}

\subsection{Spectral lags of the rest-frame spectral parameters varying from hard to soft}

Many investigations of the spectra of GRB show that the spectral
softening is an universal phenomenon. In addition, Koceviski and
Liang (2003) pointed out as the $E_{p}$ of $\nu F_{\nu}$ spectra
decays through the four BATSE channels and produce what we measure
as lag. Schaefer (2004) found that only the evolution of $E_{p}$ can
produce observed lag using the general Liang-Kargatis relation
(which describes how the peak photon energy in the spectrum changes
with time). Therefore, we first investigate the case of $E_{p}$ in
the rest-frame varying from hard to soft. Following Qin et al.
(2005), we assume a simple evolution of the peak energy $E_{0,p}$,
which follows $\log
E_{0,p}=0.1-k(\tau_{\theta}-\tau_{\theta,1})/(\tau_{\theta,2}-\tau_{\theta,1})
$ $(keV)$ for $\tau_{\theta,1}\leq\tau_{\theta}\leq\tau_{\theta,2}$.
For $\tau_{\theta}<\tau_{\theta,1}$, $\log E_{0,p}=0.1$ $(keV)$,
while for $\tau_{\theta}>\tau_{\theta,2}$, $\log E_{0,p}=0.1-k$
$(keV)$.

Here we also take $\Gamma$ = 100, 200, 300, 400, 500, 600, 700, 800,
900, 1000, respectively and $\alpha_{0}=-1$, $\beta_{0}=-2.25$,
$\sigma=1.0$ are assigned. For each $\Gamma$ we assume $k=$ 0.1,
0.2, 0.3, 0.4, 0.5, 0.6, 0.7, 0.8, 0.9, 1.0, 1.1, 1.2, 1.3 and 1.4
(they correspond to different rates of decreasing) to investigate
the spectral lag's dependence.

The local Gaussian pulse (see equation (2)) is employed to study
this issue. We adopt $\tau_{\theta,1}=8\sigma+\tau_{\theta,min}$,
$\tau_{\theta,2}=12\sigma+\tau_{\theta,min}$ and
$\tau_{\theta,0}=10\sigma+\tau_{\theta,min}$.

In this way, we can compute the spectral lags. It is found that all
of the lags are positive. A example light curve is showed in Fig. 1
(panel (b)), which illustrates the connection between the spectral
evolution and the spectral lag. This indicates the maximum of the
photon count rates at the higher energy arrives earlier than at the
lower energy in the case of hard-to-soft. Therefore, we can deduce
that only hard-to-soft spectral evolution can produce the positive
spectral lags.

\subsection{Spectral lags of the rest-frame radiation form varying from
soft to hard with time}

Band (1997) found that a small fraction of GRB spectra ($\sim$ 10\%)
vary from soft to hard with time. Kargatis et al. (1994) also found
this case. Therefore, we also investigate the case of the rest-frame
radiation form varying from soft to hard. We also assume a simple
case that the evolution of peak energy $E_{0,p}$ follows: $\log
E_{0,p}=-0.8+k(\tau_{\theta}-\tau_{\theta,1})/(\tau_{\theta,2}-\tau_{\theta,1})
$ $(keV)$ for $\tau_{\theta,1}\leq\tau_{\theta}\leq\tau_{\theta,2}$.
For $\tau_{\theta}<\tau_{\theta,1}$, and $\log E_{0,p}=-0.8$
$(keV)$, while for $\tau_{\theta}>\tau_{\theta,2}$, $\log
E_{0,p}=-0.8+k$ $(keV)$. The $\Gamma$, $k$, $\alpha_{0}$ and
$\beta_{0}$ are taken the same values as the case of hard-to-soft.
We also adopt $\tau_{\theta,1}= 8 \sigma+\tau_{\theta,min}$,
$\tau_{\theta,2}=12\sigma+\tau_{\theta,min}$ and
$\tau_{\theta,0}=10\sigma+\tau_{\theta,min}$.

According to our computation the soft-to-hard spectral evolution can
result in the negative spectral lags alone. Fig. 1 (panel (c)) also
demonstrates two example light curves indicating the connection
between the spectral evolution and the spectral lag in the case of
the rest-frame radiation form varying from soft to hard with time.

\subsection{Spectral lags of the rest-frame radiation form varying from soft to hard then to soft}
As several authors pointed out that the evolution of the low-energy
index $\alpha$ and $E_{p}$ exhibit the ``tracking'' (the observed
spectral parameters follow the same pattern as the flux or count
rate time profile, i.e. the evolution of spectral parameters exhibit
soft-to-hard-to-soft) behaviors (e.g., Ford et al. 1995; Crider et
al. 1997; and Kaneko et al. 2006; Peng et al. 2009a, 2009b). So we
investigate the case of $E_{0,p}$ varying from soft to hard then to
soft. In order to ensure that the three parameters are in the range
of observed values we adopted the following evolutionary form: $\log
E_{0,p}=-0.8+k1(\tau_{\theta}-\tau_{\theta,1})/(\tau_{\theta,0}-\tau_{\theta,1})
$ $(keV)$ for $\tau_{\theta,1}\leq\tau_{\theta}\leq\tau_{\theta,0}$,
$\log
E_{0,p}=-0.8+k1-k2(\tau_{\theta}-\tau_{\theta,0})/(\tau_{\theta,2}-\tau_{\theta,0})
$ $(keV)$ for $\tau_{\theta,0}\leq\tau_{\theta}\leq\tau_{\theta,2}$,
$\log E_{0,p}=-0.8+k1-k2$ $(keV)$ for
$\tau_{\theta}>\tau_{\theta,2}$, while for
$\tau_{\theta}<\tau_{\theta,1}$, $\log E_{0,p}=-0.8$ $(keV)$. where
the k1 and k2 denote the increasing rate of soft-to-hard phase and
the decreasing rate of hard-to-soft phase, respectively. We take
$\Gamma=$ 100, 200, 300, 400, 500, 600, 700, 800, 900, 1000,
respectively and $k1=$ 0.3, 0.4, 0.5, 0.6, 0.7, 0.8, 0.9, 1.0, and
1.1, $k2=$ 0.1, 0.2, 0.3, 0.4, 0.5, 0.6, 0.7, and 0.8, respectively.
We also adopt $\tau_{\theta,1}=8\sigma+\tau_{\theta,min}$,
$\tau_{\theta,2}=12\sigma+\tau_{\theta,min}$ and
$\tau_{\theta,0}=10\sigma+\tau_{\theta,min}$.

We use this form to investigate the spectral lags and find that the
negative lag disappears. We explore $E_{0,p}$ and $\alpha_{0}$
varying from soft to hard then to soft. The same form as $E_{0,p}$
is also adopted for $\alpha_{0}$. We find there are negative
spectral lags in this case but it does not occur in the small
$\Gamma=$100. When the $\Gamma$ is larger than 100, the negative
lags will come into being only in the case of the differences
between k1 and k2 coming to at least 0.5. That is if the increasing
rate of soft to hard is larger than the decreasing rate of hard to
soft by at least 0.5 the negative lags would occur. As the
increasing of $\Gamma$ the difference needed would decrease. The
above assumption is reasonable for the tracking pulse because it is
consistent with the result of the hardness evolutionary
characteristics of the tracking pulses investigated by Peng et al.
(2009a) that the duration of the rise phase are much shorter than
that of the decay phase. Fig. 2 illustrates four example light
curves.

\begin{figure*}
\centering \resizebox{3in}{!}{\includegraphics{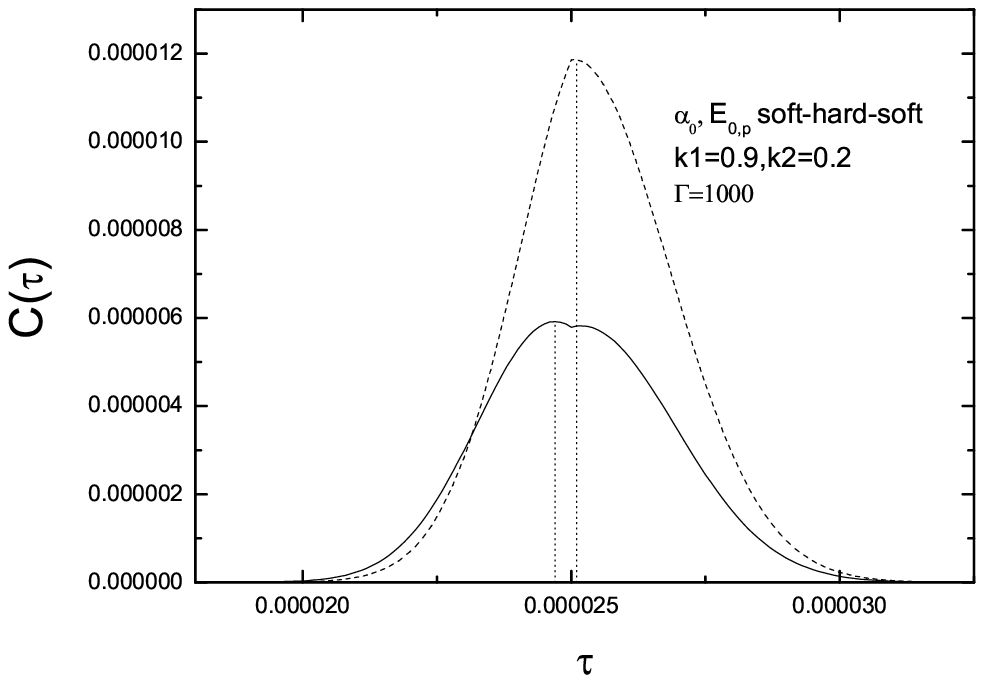}}
\resizebox{3in}{!}{\includegraphics{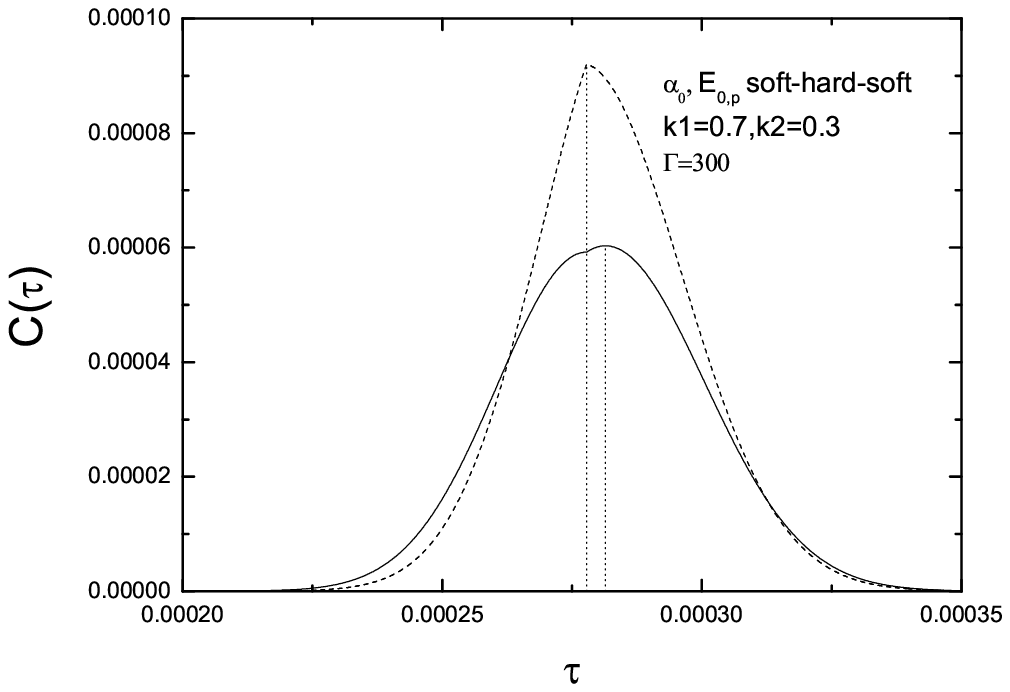}} \caption{Example light
curves of the negative (left panel) and positive (right panel)
spectral lags for the case of soft-to-hard-to-soft, where the
symbols are the same as those adopted in Fig. 1. } \label{}
\end{figure*}

\subsection{Spectral lags of a suddenly shining and gradually dimming intrinsic emission profile}
If the different intrinsic emission profiles are different in
producing the spectral lags. Similar to Qin et al. (2004) and Shen
et al. (2005) we also consider the one-sided exponential decay and
power-law decay profile, which are listed as follows, respectively:

\begin{equation} \widetilde{I}(\tau _\theta
)=I_0\exp [-(\frac{\tau _\theta - \tau _{\theta,\min}}\sigma )]
\qquad (\tau _{\theta ,\min }\leq \tau _\theta ),
\end{equation}

\begin{equation} \widetilde{I}(\tau _\theta
)=I_0(1-\frac{\tau _\theta - \tau _{\theta,\min}}{
\tau_{\theta,\max}-\tau_{\theta,\min }})^{\mu} \qquad (\tau _{\theta
,\min }\leq \tau _\theta ),
\end{equation}

We assume the $\tau _{\theta ,\min}$ = 0 and $\mu$ = 2 in the
following analysis. The hard-to-soft and soft-to-hard spectral
evolution are investigated to check if the two sorts of local pulses
also result in the same results as those of the Gaussian pulses. The
same evolutional patterns adopted above are employed to explore this
issue. We find the results are in agreement with that of Gaussian
local pulses. The example plots in Figs. 3 and 4 also illustrate the
connection between the spectral evolution and spectral lag for the
exponential decay and power-law decay, respectively. This suggests
the intrinsic emission profile has no effect on the connection
between the spectral evolution and the spectral lags. Likewise, the
hard-to-soft evolution produces positive lags and the soft-to-hard
evolution leads to negative lags if we adopt other rest-frame local
pulses.

\begin{figure*}
\centering \resizebox{3in}{!}{\includegraphics{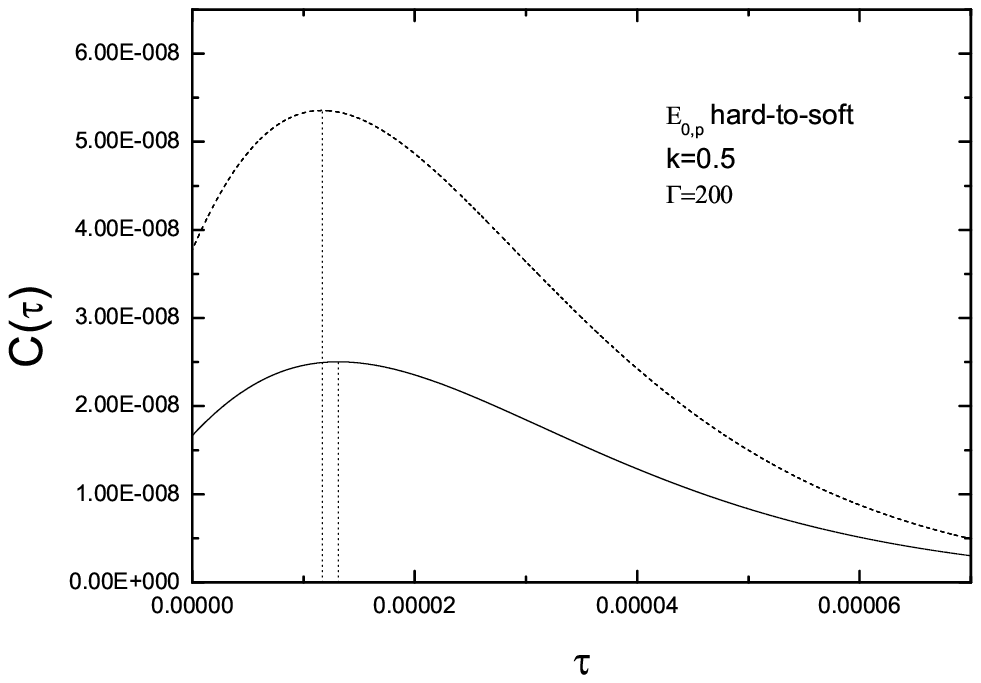}}
\resizebox{3in}{!}{\includegraphics{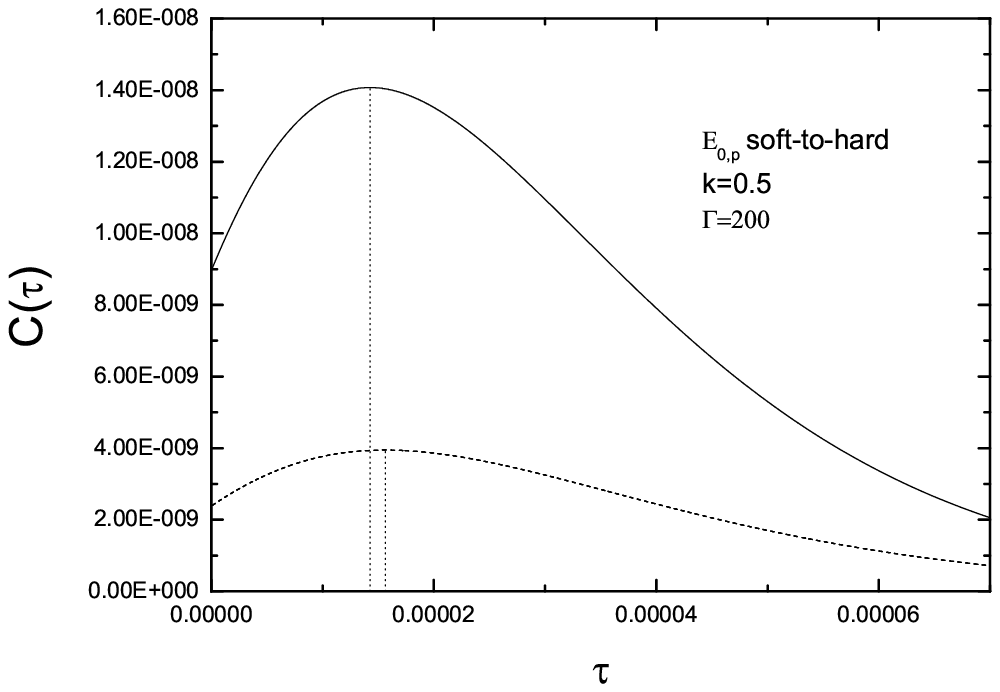}} \caption{Example light
curve of $E_{0,p}$ varies from hard to soft (left panel) and soft to
hard (right panel) for the case of the exponential decay local
pulse, where the symbols are the same as those adopted in Fig. 1.}
\label{}
\end{figure*}

\begin{figure*}
\centering \resizebox{3in}{!}{\includegraphics{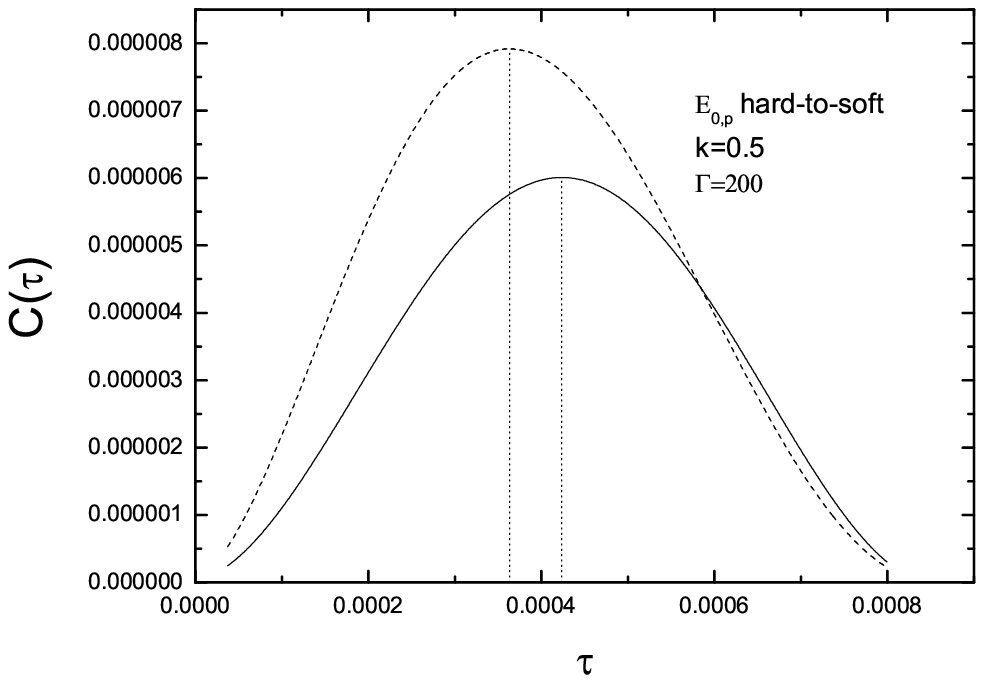}}
\resizebox{3in}{!}{\includegraphics{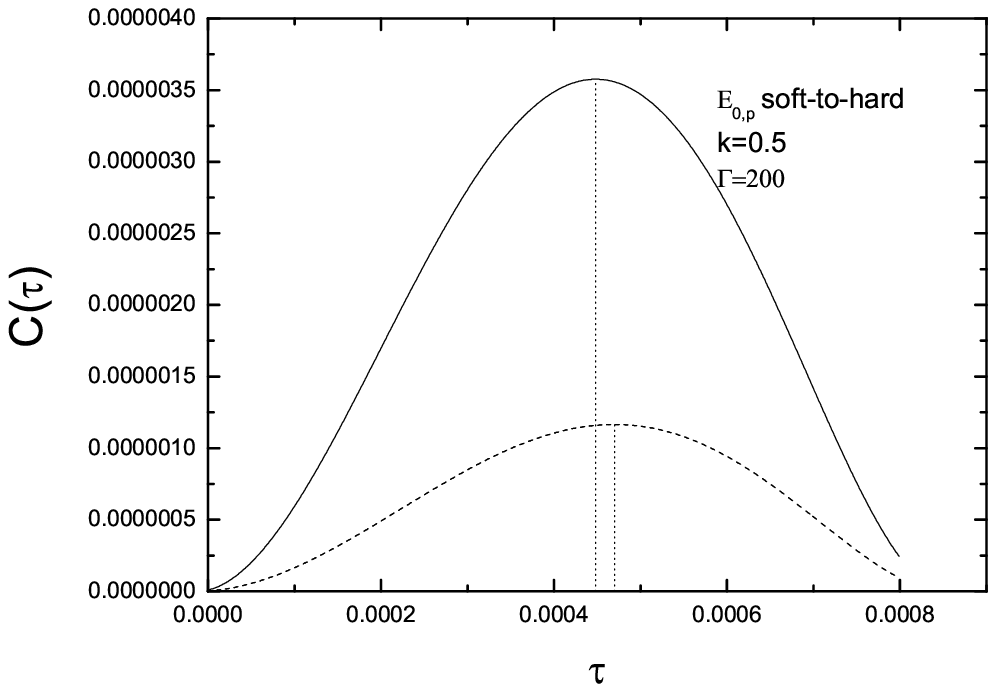}} \caption{Example light
curve of $E_{0,p}$ varies from hard to soft (left panel) and soft to
hard (right panel) for the case of power law decay local pulses,
where the symbols are the same as those adopted in Fig. 1.} \label{}
\end{figure*}

\section{Spectral lags caused by the spectral evolution and the curvature effect}
As previous section pointed out that the curvature effect must be
play a role on the producing of temporal and spectral profile we
also take the effect into account to study the spectral lag. The
evolutionary form, hard-to-soft and soft-to-hard, are combined with
curvature effect to explore the influence of the combined effects on
the spectral lags. That is we consider the radiation emitted from a
big area with $\theta_{min}=0$ and $\theta_{max} =\pi/2$ where the
photons emitted travel different distances to the observer. The
photons on the line of sight arrive at us earlier than that of off
sightline. In this way the lags due to curvature effect is always
positive and the hard-to-soft evolution also make the lag is
positive. Therefore we think the spectral lags caused by the
combined effect must be increase.

We first consider the case of $E_{0,p}$ varying from hard to soft
combined with the curvature effect. In addition, we also assume the
evolution of $E_{0,p}$ follows the same means as the above section.
To compare the lags produced by the spectral evolution alone with
that by the spectral evolution and curvature effect together we plot
two sorts of lags together in Fig. 5 (panel (a)) by considering a
set of parameter values. Note that the histogram plots may mean
nothing because when one consider another set of parameter values
they get a different result. It is shown that (i) the average value
of spectral lags caused by spectral evolution and combined effect
are 0.1086 s and 0.1692 s, respectively and (ii) the corresponding
medians are 0.007 s and 0.016 s, respectively. These show the lags
caused by the combined effect increase indeed. Whether the curvature
effect also make the lags increase in the case of soft-to-hard or
not? We also check the spectral lags resulting from the curvature
effect and evolution of $E_{0,p}$ following soft-to-hard. The lags
produced by spectral evolution alone and the combined effect are
also compared. The mean values are -0.1200 s and -0.0114 s for
spectral evolution and combined effect, respectively. Whereas the
corresponding median values are -0.0102 s and -0.0005 s,
respectively. Fig. 5 (panel (b)) also shows that the lags indeed
increase in the case of soft-to-hard. In addition, we also find due
to the curvature effect some negative lags change into positive
ones. In view of the analysis of the above two cases we can conclude
that the curvature effect indeed make the lags increase
significantly.
\begin{figure*}
\centering \resizebox{3in}{!}{\includegraphics{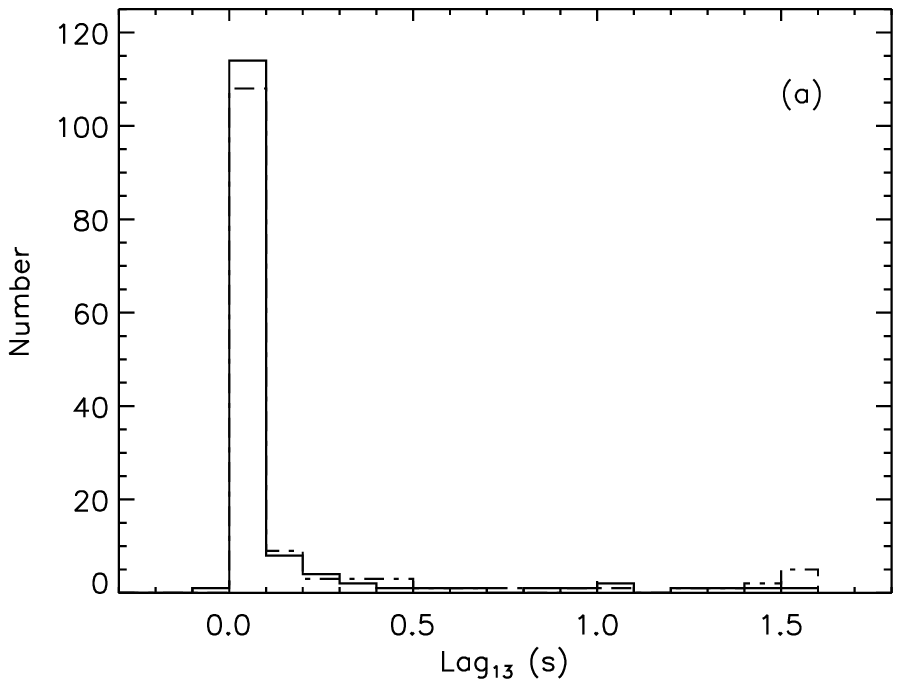}}
\resizebox{3in}{!}{\includegraphics{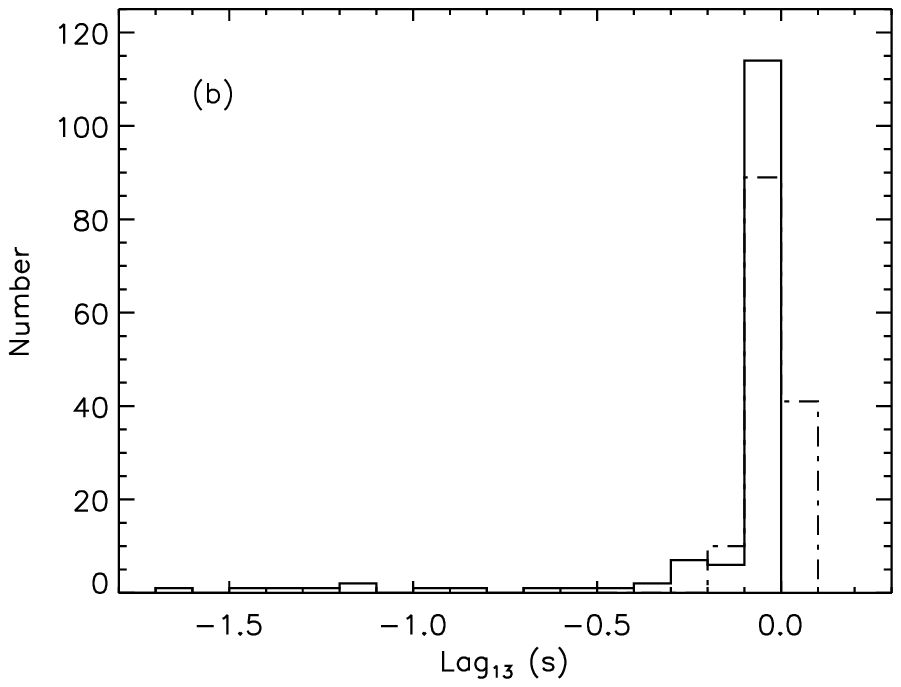}} \caption{The
comparisons of lag distribution between the case of hard-to-soft
(panel a) and soft-to-hard (panel b) evolution alone (solid lines)
and the case of spectral evolution and curvature effect
simultaneously (the dashed lines).} \label{}
\end{figure*}

\section{Observed spectral lags of the hard-to-soft pulses}

The above analysis shows that the spectral lags resulting from the
hard-to-soft spectral evolution are positive and the negative lags
may come from the soft-to-hard or soft-to-hard-to-soft spectral
evolution. We wonder if the observations are consistent with the
predictions. The pulses exhibit hard-to-soft, soft-to-hard and
soft-to-hard-to-soft spectral evolution should be checked. However,
the spectral parameter evolution of observed pulses generally
exhibit hard-to-soft and ``tracking'' (i.e. soft-to-hard-to-soft).
Both behaviors are defined by $E_{p}$ and/or low-energy index
evolution, which are found by several authors (Kaneko et al. 2006;
Ford et al. 1995; Crider et al. 1997). With ordinary Poisson noise
in GRB light curves, there will be significant scatter in the
measured lags. The percentage of lags increase substantially towards
small values, with many (the high luminosity events) having
near-zero lags. In such a case, we expect there to be many
``measured'' negative lags, even if all lags are positive-definite.
That is, a convolution of a positive-definite lag distribution with
the known measurement errors apparently reproduces the observed lag
distribution including the measured negative lags. With this, a good
case can be made that negative lags do not exist. The existence of
multiple overlapping pulses could easily produce negative lags, even
though the lags for each pulse are positive. All it takes is two
nearly overlapping pulses with the second one being harder than the
first. In all, we are not convinced in the significant existence of
negative lags.

Therefore, we select a sample only containing the FRED (fast rise
and exponential decay) pulses provided by Kocevski et al. (2003) to
check the theoretical predication since these bursts exhibit clean,
single-peaked or well-separated in multi-peaked events. We can
obtain directly the spectral parameters of bright bursts from Kaneko
et al. (2006). The weaker bursts are also analyzed using RMFIT
software and fitted with Band model. Similar to Peng et al. (2009a,
2009b, 2010) and Ryde \& Svensson (2002) a signal-to-noise ratio
(S/N) of the observations of at least $\geq$ 30 to get higher time
resolution are adopted. For the hard-to-soft pulses only that
containing at least 1 data points of $E_{p}$ before the peak time of
pulse and 4 in all are included in the our analysis. Finally, we
select 15 hard-to-soft pulses as our sample to test our predictions.
Illustrated in Fig. 6 are the spectral and temporal behaviors for
our sample.

\begin{figure*}
\centering
\resizebox{2.2in}{!}{\includegraphics{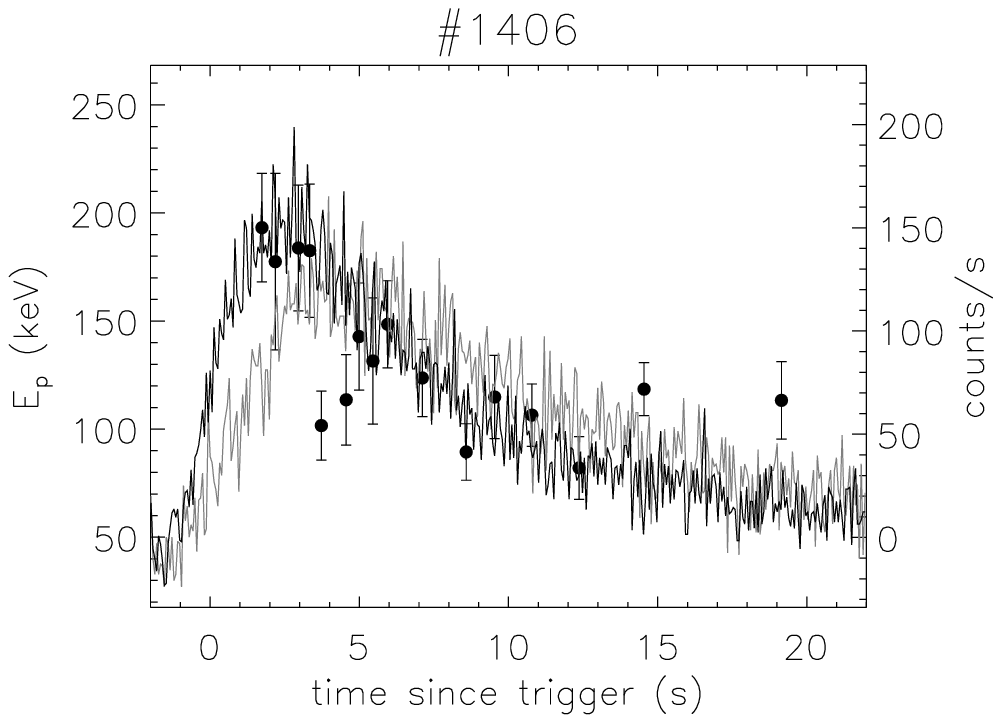}}
\resizebox{2.2in}{!}{\includegraphics{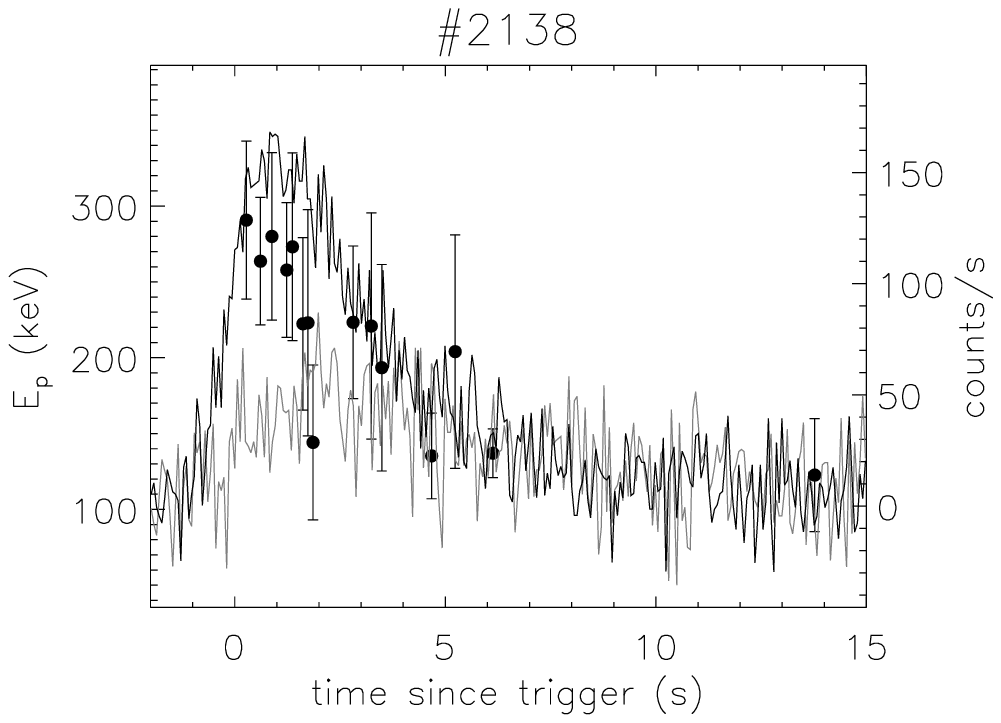}}
\resizebox{2.2in}{!}{\includegraphics{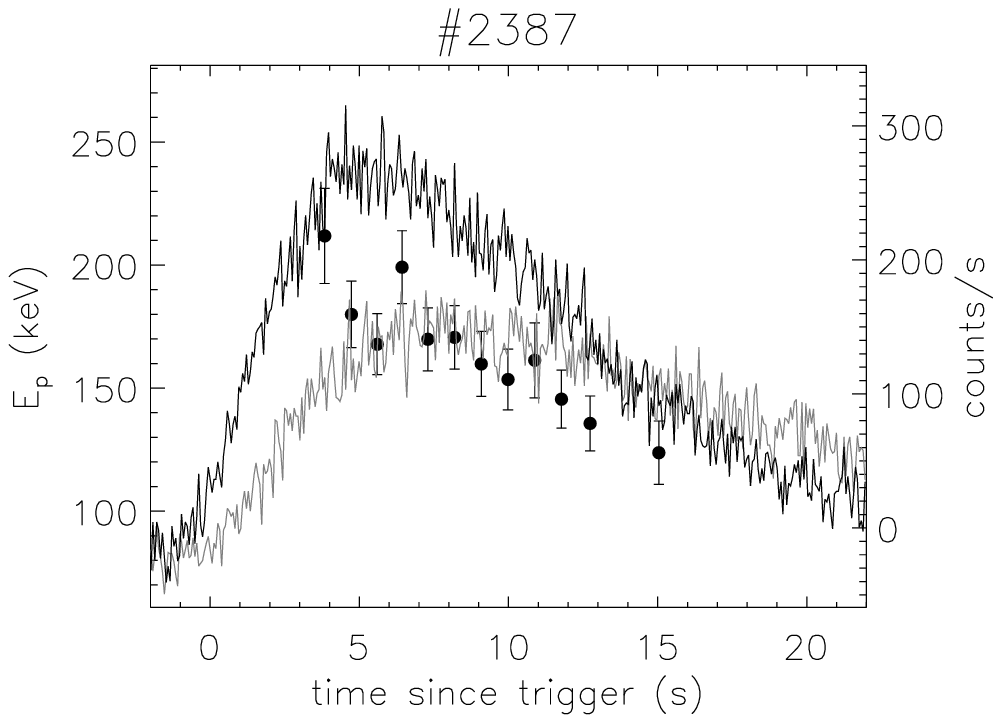}}
\resizebox{2.2in}{!}{\includegraphics{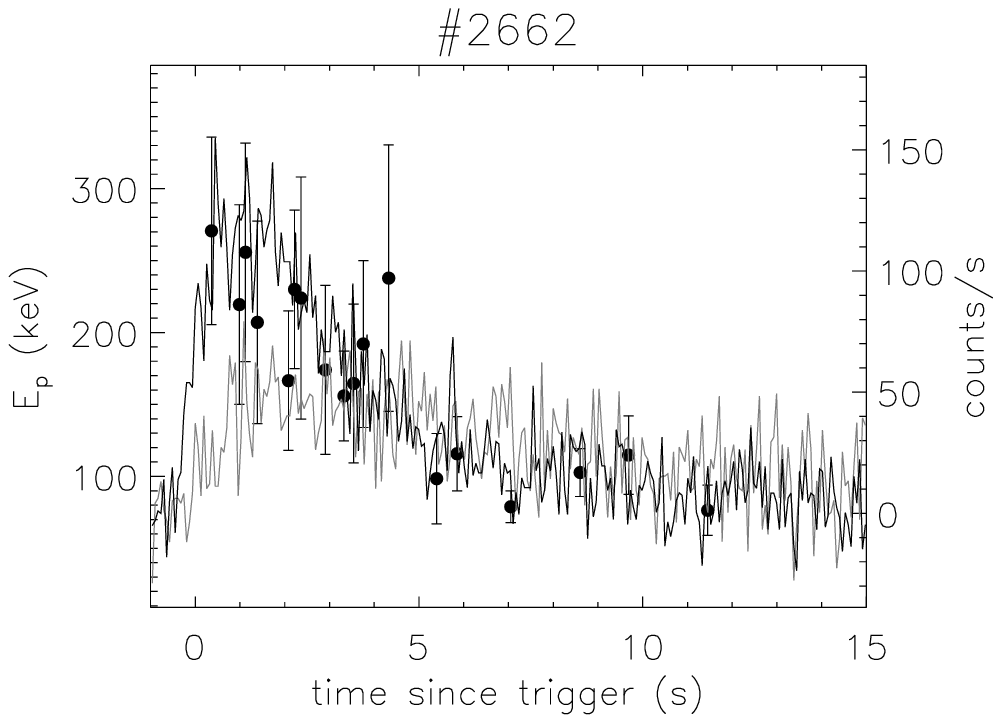}}
\resizebox{2.2in}{!}{\includegraphics{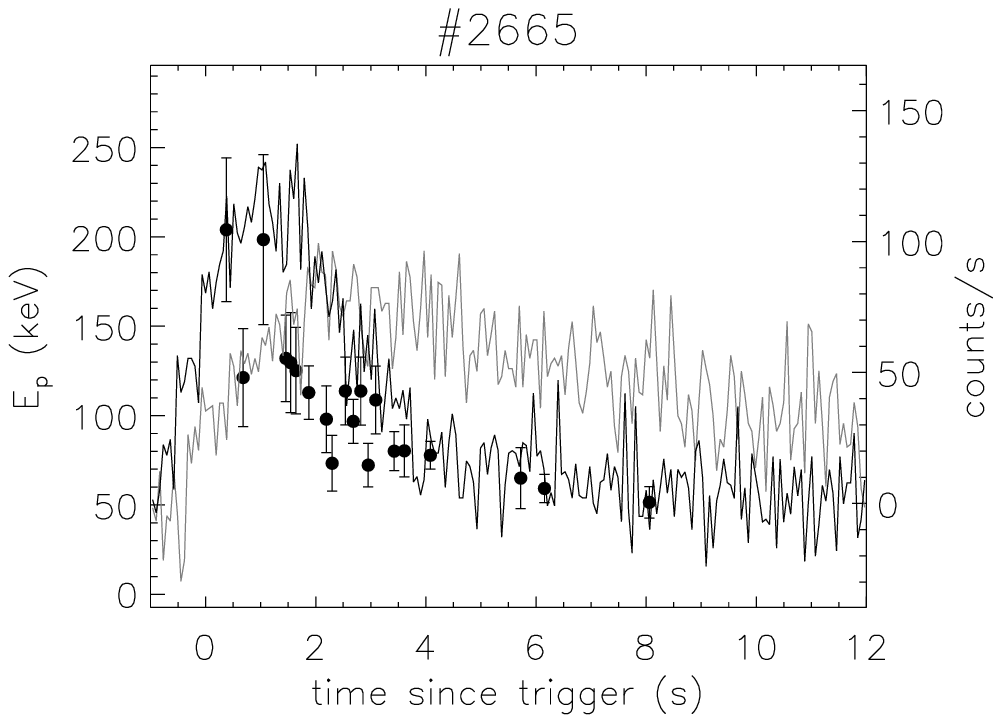}}
\resizebox{2.2in}{!}{\includegraphics{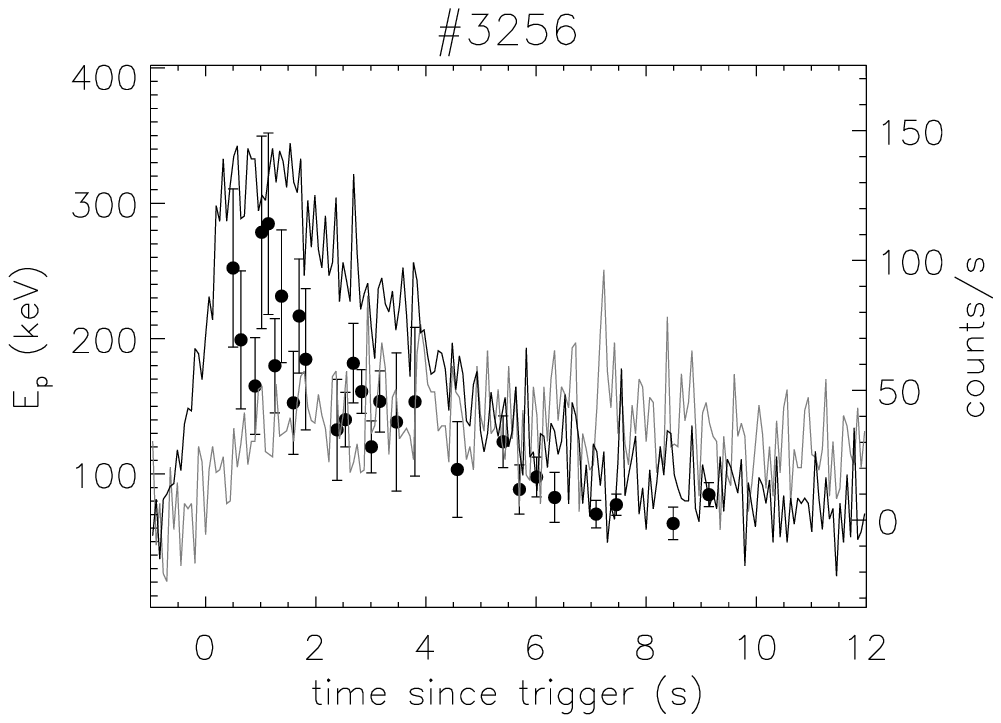}}
\resizebox{2.2in}{!}{\includegraphics{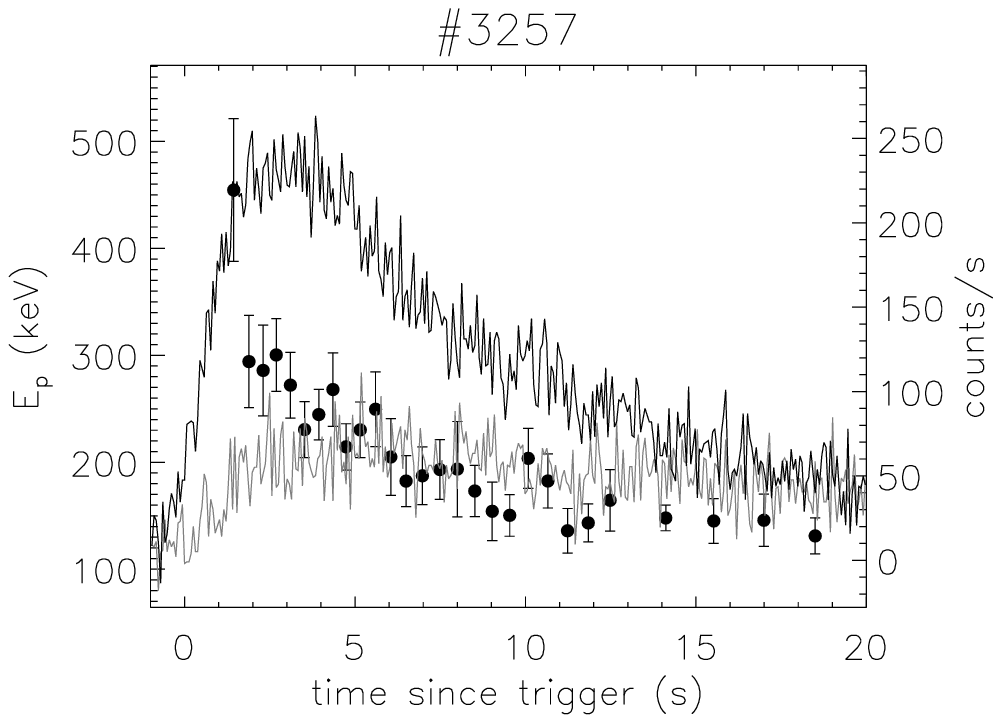}}
\resizebox{2.2in}{!}{\includegraphics{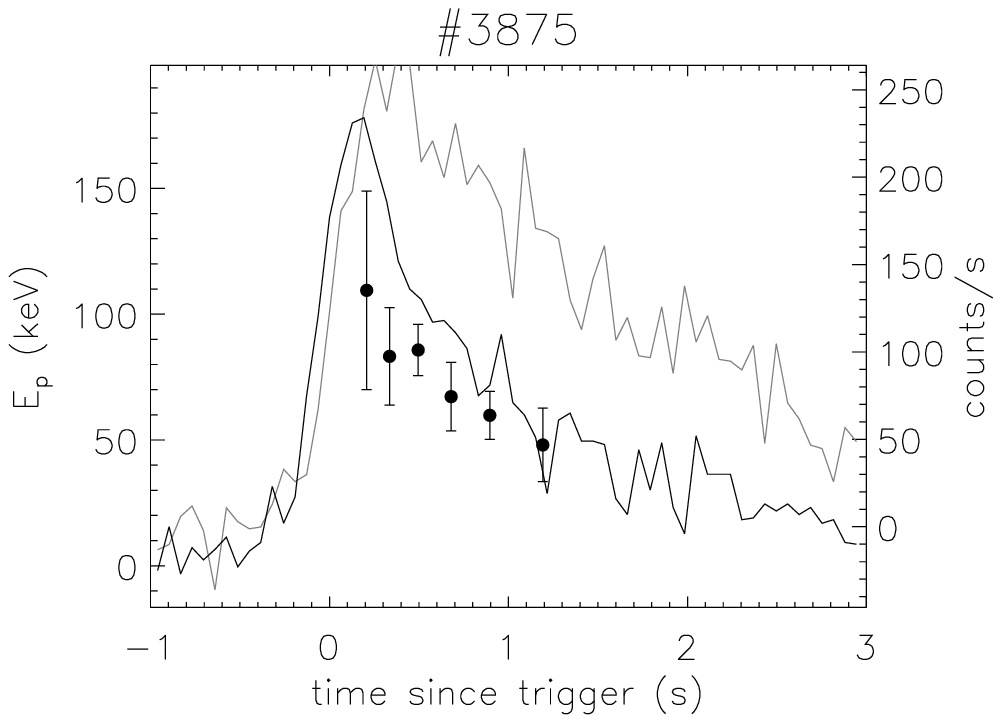}}
\resizebox{2.2in}{!}{\includegraphics{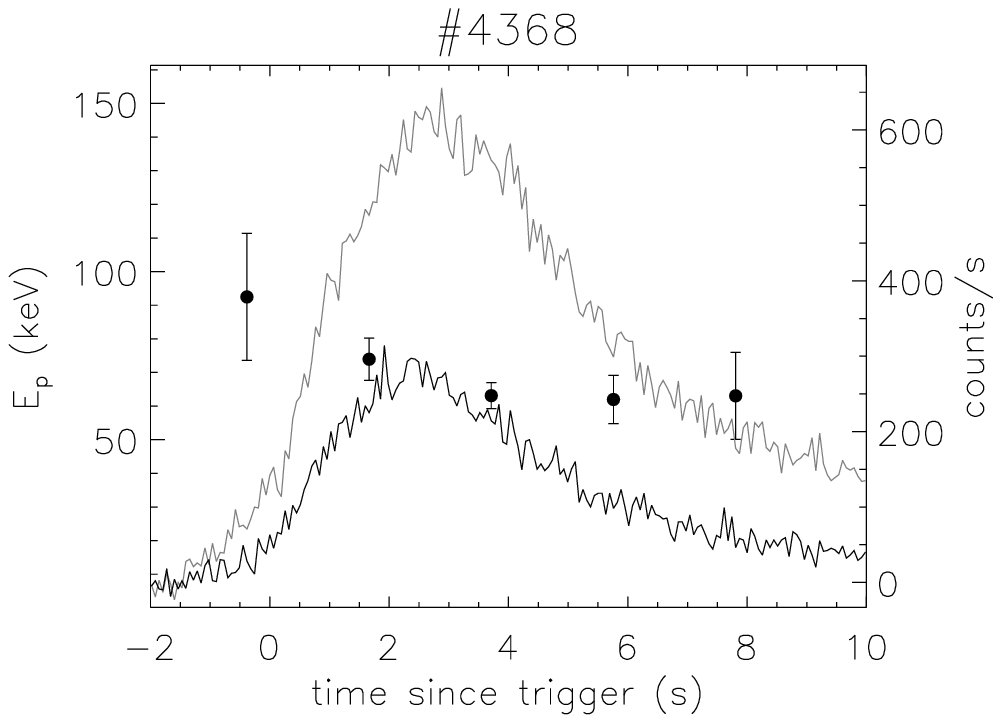}}
\resizebox{2.2in}{!}{\includegraphics{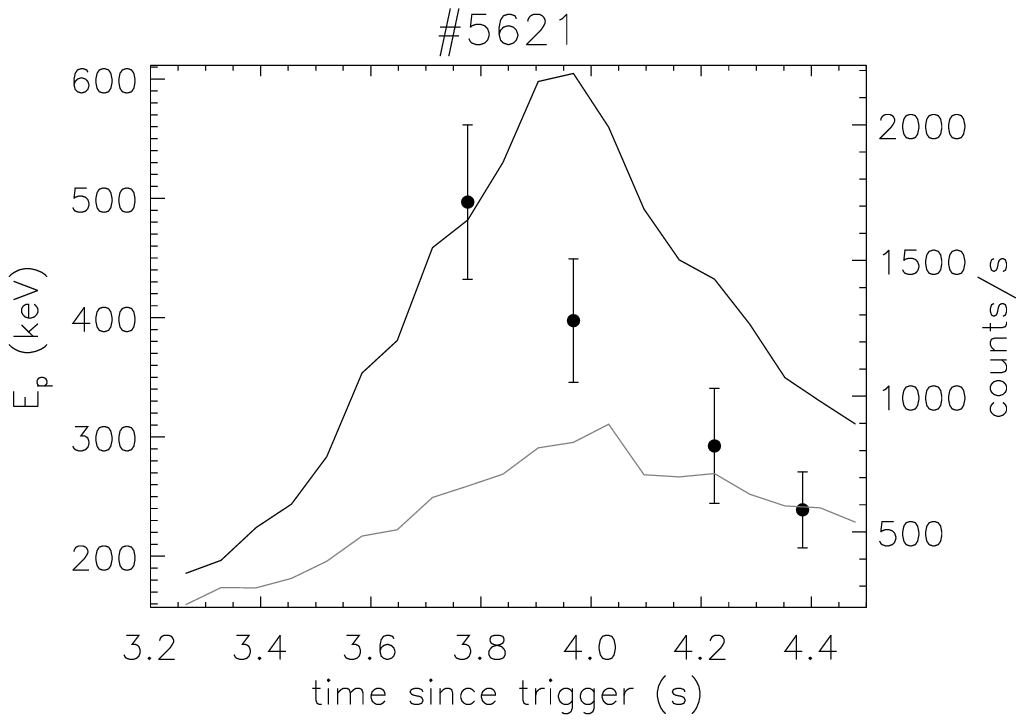}}
\resizebox{2.2in}{!}{\includegraphics{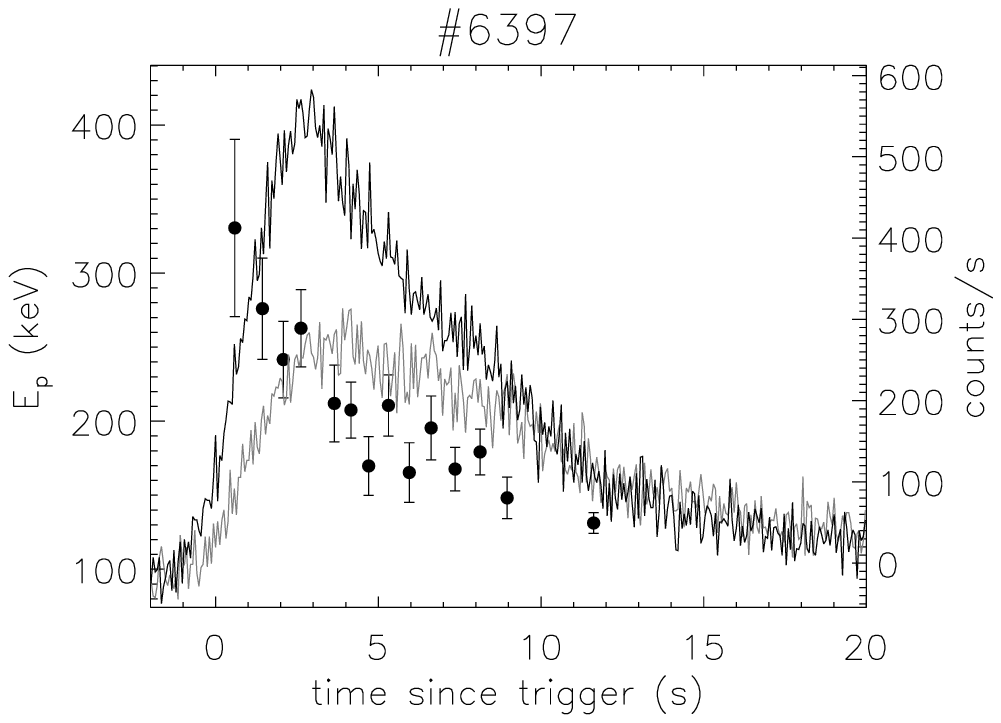}}
\resizebox{2.2in}{!}{\includegraphics{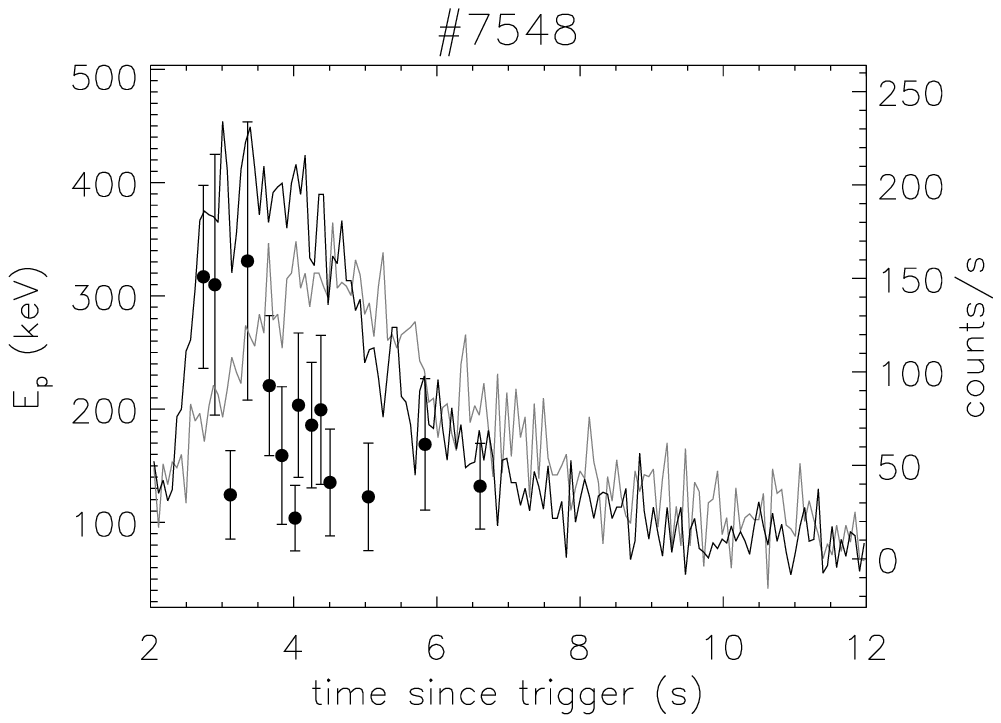}}
\resizebox{2.2in}{!}{\includegraphics{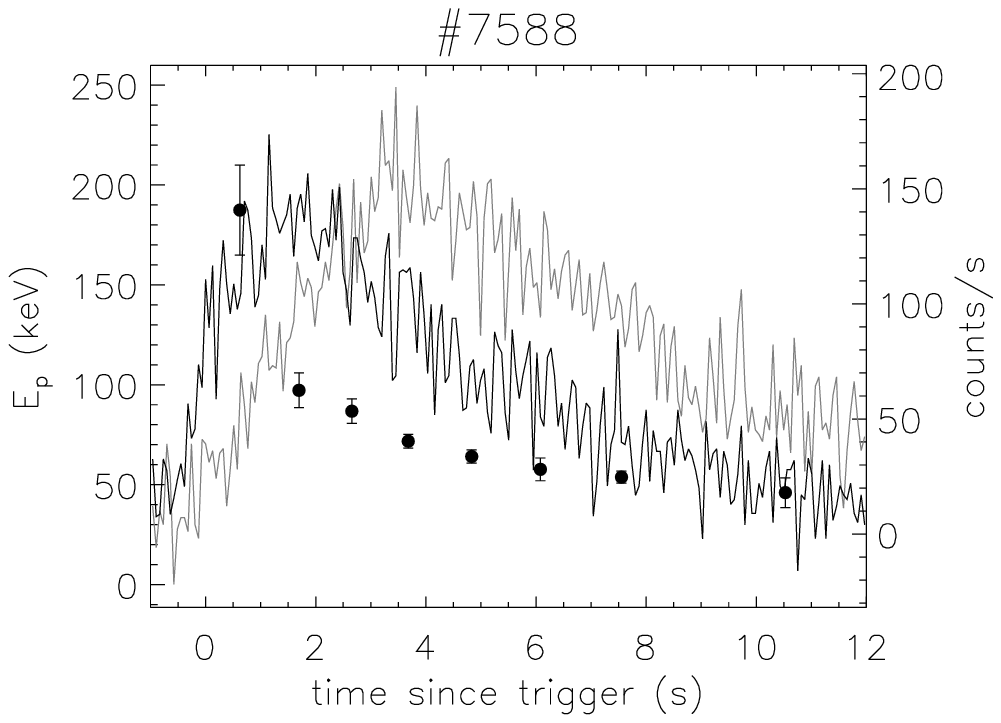}}
\resizebox{2.2in}{!}{\includegraphics{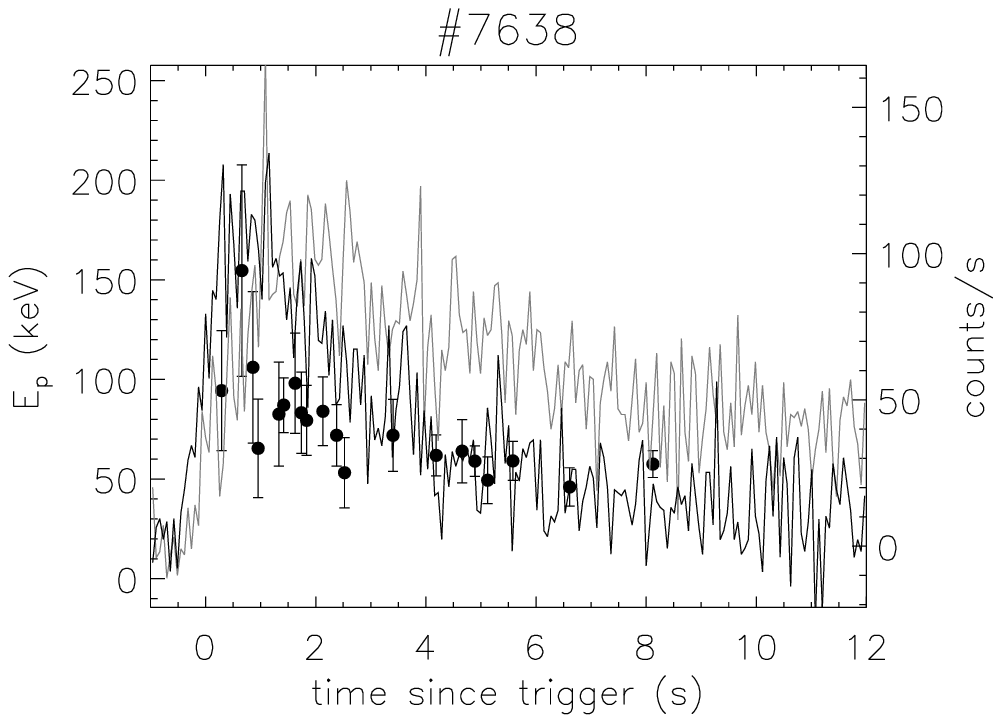}}
\resizebox{2.2in}{!}{\includegraphics{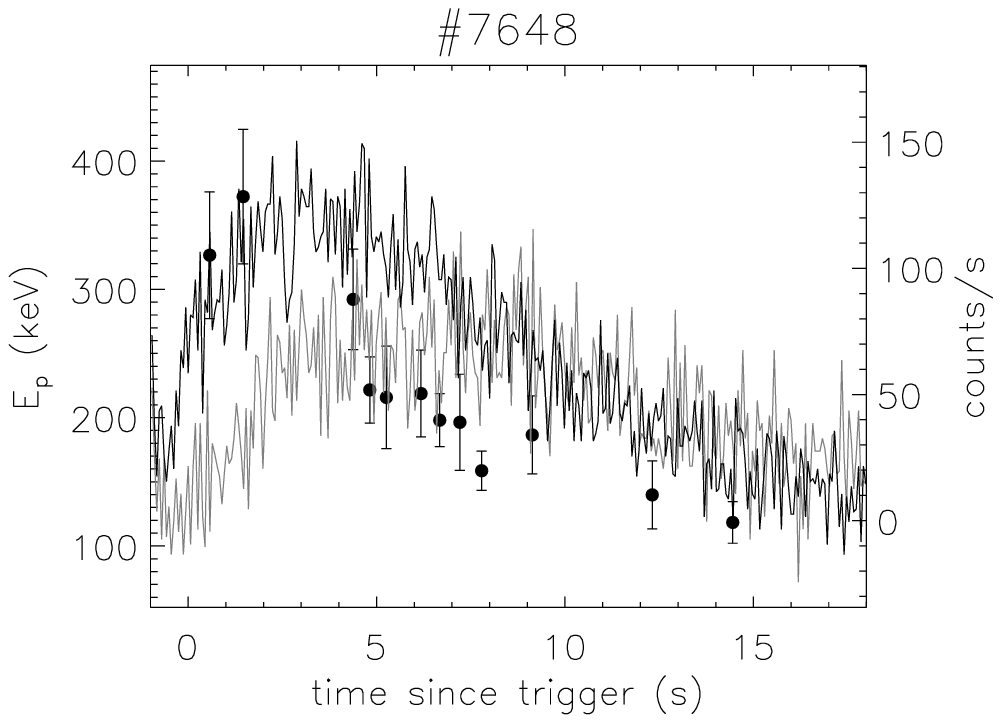}}

\caption{Spectral evolution of $E_{p}$ and temporal behaviors of GRB
pulses of BATSE channels 3 and 1 in our sample for the case of
hard-to-soft, where the black lines represent the light curves of
channel 3. The positive lags are evident.} \label{}
\end{figure*}

We first employ the widely used cross-correlation function (CCF) to
compute the spectral lags between the pulses of the same burst seen
in the BATSE channels 1 and 3. Similar to Norris et al. (2000) we
adopt a cubic polynomial to fit the peak of the resulting discrete
CCF function and take the peak of the cubic polynomial form as the
lag. The error in the average CCF lag, $lag_{13}$, is found by
simulations with Gaussian noise added to each of the two channels.
These spectral lags resulting from the evolution of hard-to-soft are
listed in Table 1. We find no negative lags in the hard-to-soft
pulses.

Recently Hakkila et al. (2008) point out that the CCF and pulse peak
lags occasionally disagree. In order to further test it we also find
pulse peak lags which are defined as the differences between the
pulse peak in BATSE channels 1 and 3. In order to find the peak time
of pulse we must select a pulse model to fit these pulses. Following
Peng et al. (2006) we only select the function presented in equation
(22) of Kocevski et al. (2003). We fit those selected
background-subtracted pulses using the KRL function. Our fittings
are constrained on the channels 1 and 3 to find their difference of
peak times. The best fitting parameters are depended on which
$\chi_{\nu}^{2}$ of the fitting model is smaller. The pulse peak
lags along with the fitting $\chi_{\nu}^{2}$ in channels 1 and 3 are
also listed in Table 1. From Table 1 we find all the lags of
hard-to-soft pulse are positive. We also check the CCF lags and peak
lags by smoothing the light curves with the DB3 wavelet (Qin et al.
2004) and obtain the same results (see, Table 1).

\begin{table*}
\centering \caption{A list of various observed lags of GRB pulses
whose spectral evolution following from hard to soft}
\begin{tabular}{ccccccccccc}
\hline trigger & st (s) & et (s) & $lag_{13,p}$ (s) & $dlag_{13,p}$
(s) & $lag_{13,c}$ (s) & $dlag_{13,c}$ (s) & $\chi_{\nu,1}^{2}$  & $\chi_{\nu,3}^{2}$ \\
\hline
1406 & -2.0 & 15.0  & 1.864 $\pm$ 0.169 &1.857 $\pm$ 0.169 & 1.152 $\pm$ 0.019 & 0.487 $\pm$ 0.509 & 1.05 & 1.07 \\
2138 & -2.0 & 15.0  & 0.988 $\pm$ 0.302 &1.001 $\pm$ 0.316 & 1.024 $\pm$ 0.057 & 0.604 $\pm$ 0.178 & 1.10 & 0.99 \\
2387 & -2.0 & 20.0  & 2.188 $\pm$ 0.204 &2.169 $\pm$ 0.207 & 1.856 $\pm$ 0.019 & 1.574 $\pm$ 0.177 & 1.12 & 1.15 \\
2662 & -2.0 & 15.0  & 1.331 $\pm$ 0.285 &1.355 $\pm$ 0.272 & 0.768 $\pm$ 0.159 & 0.712 $\pm$ 0.139 & 1.01 & 1.17 \\
2665 & -2.0 & 15.0  & 1.552 $\pm$ 0.196 &1.555 $\pm$ 0.196 & 1.507 $\pm$ 0.503 & 1.417 $\pm$ 0.148 & 0.89 & 0.91 \\
3256 & -2.0 & 15.0  & 2.120 $\pm$ 0.428 &2.828 $\pm$ 0.390 & 1.532 $\pm$ 0.806 & 1.325 $\pm$ 0.147 & 1.04 & 1.07 \\
3257 & -1.0 & 15.0  & 1.667 $\pm$ 0.517 &1.606 $\pm$ 0.509 & 1.283 $\pm$ 0.075 & 1.134 $\pm$ 0.094 & 0.97 & 0.92 \\
3875 & -1.0 & 5.0   & 0.262 $\pm$ 0.029 &0.257 $\pm$ 0.031 & 0.162 $\pm$ 0.026 & 0.169 $\pm$ 0.011 & 1.40 & 1.18 \\
4368 & -2.0 & 10.0  & 0.346 $\pm$ 0.047 &0.346 $\pm$ 0.048 & 0.265 $\pm$ 0.027 & 0.265 $\pm$ 0.013 & 1.03 & 0.82 \\
5621 & 3.2  & 4.5   & 0.032 $\pm$ 0.022 &0.076 $\pm$ 0.011 & 0.028 $\pm$ 0.002 & 0.030 $\pm$ 0.001 & 1.24 & 1.71 \\
6397 & -2.0 & 20.0  & 1.158 $\pm$ 0.082 &1.160 $\pm$ 0.082 & 0.942 $\pm$ 0.031 & 0.934 $\pm$ 0.171 & 1.17 & 1.70 \\
7548 & 2.0  & 12.0  & 0.814 $\pm$ 0.075 &0.816 $\pm$ 0.075 & 0.596 $\pm$ 0.232 & 0.628 $\pm$ 0.063 & 0.88 & 1.10 \\
7588 & -2.0 & 15.0  & 2.230 $\pm$ 0.120 &2.220 $\pm$ 0.120 & 1.125 $\pm$ 0.292 & 1.427 $\pm$ 0.390 & 0.87 & 0.93 \\
7638 & -2.0 & 15.0  & 1.124 $\pm$ 0.114 &1.420 $\pm$ 0.134 & 0.968 $\pm$ 0.012 & 0.931 $\pm$ 0.162 & 1.15 & 1.04 \\
7648 & -2.0 & 15.0  & 2.490 $\pm$ 0.243 &2.501 $\pm$ 0.248 & 1.574 $\pm$ 1.202 & 1.528 $\pm$ 0.386 & 1.01 & 1.03 \\
\hline
\end{tabular}

Note: st and et denotes the start and end time since bursts trigger
of selected pulses, respectively. $lag_{13,p}$ and $lag_{13,c}$
represents peak time lag and CCF lag, while $dlag_{13,p}$ and
$dlag_{13,c}$ are the corresponding lags but the light curve having
been smoothed with the DB3 wavelet, $\chi_{\nu,1}^{2}$ and
$\chi_{\nu,3}^{2}$ is the fitting $\chi_{\nu}^{2}$ of channels 1 and
3, respectively.
\end{table*}

\section{Conclusions and discussions}

Spectral evolution is an established characteristic in GRBs.
Moreover the trend for their high-energy photons to arrive before
the lower-energy ones (``hard-to-soft'' evolution) is universal
phenomenon, which leads to positive lags. Kocevski \& Liang (2003)
showed the robust connection between the spectral lag measured in
GRBs and the hard-to-soft evolution of the burst spectra by
employing a sample of clean single-peaked bursts with measured lag.
In this paper, we have investigated the connection between the
evolution of the rest-frame spectral parameters and the spectral lag
using a theoretical model derived by Qin (2002) and Qin et al.
(2004). We first assume an intrinsic Band spectrum and a Gaussian
emission profile to study the issue. Our theoretical analysis shows
that the connection exists not only between the hard-to-soft
evolution of the burst spectra and the spectral lag but between the
soft-to-hard evolution and the spectral lag as well as between the
soft-to-hard-to-soft and the spectral lag. In addition, the
hard-to-soft spectral parameters evolution can produce the positive
lags, while the negative lags may be resulting from the evolution of
soft-to-hard as well as soft-to-hard-to-soft. We also take account
of the other local pulse forms studied by Qin et al. (2004) and find
that the spectral lags resulting from these different local pulses
are consistent with those of the Gaussian pulse.

Let us check whether there are any differences in the measurements
between peak lags, $Lag_{13}$, adopted in the above analysis and the
widely used cross correlation function lags, $CCF_{13}$ (e.g.,
Norris et al. 2000). Similar to Band (1997) and Norris et al. (2000)
we fitted a cubic polynomial to the peak of the resulting discrete
CCF function and find the difference of the corresponding peak time.
In Fig. 7 the peak lag, $Lag_{13}$, is plotted as a function of the
CCF lag, $CCF_{13}$. A linear fit is given which shows that there is
a good correspondence between the two methods. The best fit is
$Lag_{13} = (-0.03 \pm 0.03)+(1.28 \pm 0.23) \times CCF_{13}$. The
values of intercept and slope are consistent with the results of
Ryde et al. (2005). Therefore, the peak lags we used in this study
would not affect our analysis results.

\begin{figure}
\centering \resizebox{3.0in}{!}{\includegraphics{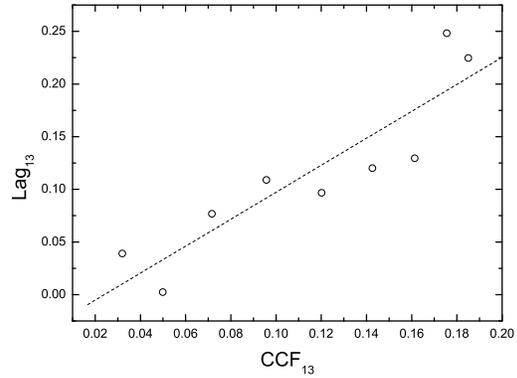}}
\caption{Peak lag (s) versus CCF lag (s). A good linear
correspondence exists, as expected.} \label{}
\end{figure}

We select a sample consisting of 15 pulses whose peak energy $E_{p}$
follow hard-to-soft evolution to test our theoretical predictions.
Our analysis shows that all of the CCF lags and analytic lags of
hard-to-soft pulses are positive within the range of uncertainty. In
order to decrease the effect of noise on our results we smoothed the
data with the DB3 wavelet (the second-class decomposition) using the
MATLAB software (Qin et al. 2004). Then we re-calculate the two
types of lags (see, Table 1) and find the lags are also positive.
This indicates hard-to-soft spectral evolution may indeed produce
positive lags, which is in good agreement with our theoretical
prediction.

Currently we can not ascertain if the negative lag predicted by our
adopted model are consistent with observation because we are not
convinced in the significant existence of negative lags in our
observation. The main reasons mentioned in the previous section are
the near-zero lags of most pulse as well as the existence of
multiple overlapping pulses. Even then we can make a theoretical
predication about the producing of the negative lags, i.e. the
soft-to-hard and the soft-to-hard-to-soft spectral evolution may
give rise to the negative lags, which reserves the further
investigations in the future observations.

When taking the spectral evolution and curvature effect into account
simultaneously we find the lags are larger than those associated
with the case that we only consider the spectral evolution (see,
Fig. 5). The two factors must play the roles simultaneously on the
producing of the spectral lags since the curvature effect must be
play a important role on the producing of spectral and temporal
profile of the GRB. In other words, the observed lags should suffer
from this two factors simultaneously. Therefore we argue that the
number of the observed negative lag should be much smaller than
positive lags. Recently, Chen et al. (2005) used the data acquired
in the time-to-spill (TTS) mode for long GRBs by BATSE and found
that positive lags is the norm in this selected sample set of long
GRBs. While relatively few in number, some pulses of several long
GRBs do show negative lags, which is also consistent with our
theoretical predication.

Both the theory and observation suggest the connection between the
spectral evolution and the spectral lags indeed exists. Therefore,
we must make clear the mechanism of intrinsic spectral evolution in
order to reveal the underlying physics of the spectral lag. To
explore this, we must first understand the mechanism that produces
breaks in the GRB spectra and what causes its evolution but it is a
difficult issue. As for the physical model of hard-to-soft spectral
evolution several authors have proposed the possible origin. For
example, Tavani (1996) showed the behavior of hard-to-soft spectral
evolution are caused by the variation of the average Lorentz factor
of pre-accelerated particles and the strength of the local magnetic
field at the GRB site as the synchrotron emission evolves within the
burst. While Liang (1997) proposed a physical model of hard-to-soft
spectral evolution in which impulsively accelerated non-thermal
leptons cool by saturated Compton upscattering of soft photons.
Kocevski \& Liang (2003) pointed out it is a harder question due to
uncertainties involved with the microphysics of GRBs. They thought
the interpretation of peak energy $E_{p}$ (and hence its evolution)
depends on the radiation mechanism that is used to explain the GRB
spectra. Moreover they also attempted to examine several mechanisms
that could produce the decay of the GRB spectra and discuss how this
evolution could be connected to the bursts¡¯ luminosity for each
model. Whereas for the soft-to-hard-to-soft spectral evolution few
attempts have been made to explain it. Peng et al. (2009a) argued
that the phenomenon may be caused by both kinematic and dynamic
process. However they did not provide further evidences to
distinguish which one is dominant. The issue of spectral evolution
of hard-to-soft and soft-to-hard-to-soft is still unclear, which
deserves careful studies and further investigations.

\acknowledgements This work was supported by the Science Fund of the
Education Department of Yunnan Province (08Y0129, 09Y0414), the
Natural Science Fund of Yunnan Province (2009ZC060M), the key
project of applied basic study of Yunnan Province (2008CC011),
National Natural Science Foundation of China (No. 10778726).


%
%
%
%
%

\end{document}